# Supporting Finite Element Analysis with a Relational Database Backend
# Part I: There is Life beyond Files

Gerd Heber
Cornell Theory Center, 638 Rhodes Hall, Cornell University, Ithaca, NY 14850, USA
heber@tc.cornell.edu

Jim Gray
Microsoft Research, San Francisco, CA, 94105, USA
Gray@Microsoft.com

# Supporting Finite Element Analysis with a Relational Database Backend

## Part I: There is Life beyond Files

Gerd Heber and Jim Gray

Cornell Theory Center, 638 Rhodes Hall, Ithaca, NY 14850, USA

heber@tc.cornell.edu

Microsoft Research, 455 Market St., San Francisco, CA 94105, USA

gray@microsoft.com

**Abstract**: *In this paper, we show how to use a Relational Database Management System in support of Finite Element Analysis. We believe it is a new way of thinking about data management in well-understood applications to prepare them for two major challenges, - size and integration (globalization). Neither extreme size nor integration (with other applications over the Web) was a design concern 30 years ago when the paradigm for FEA implementation first was formed. On the other hand, database technology has come a long way since its inception and it is past time to highlight its usefulness to the field of scientific computing and computer based engineering. This series aims to widen the list of applications for database designers and for FEA users and application developers to reap some of the benefits of database development.*

## Introduction

This is the first article of a three-part series. Together, they show how Relational Database Management Systems (RDBMS) can dramatically simplify the well established engineering tool, Finite Element Analysis (FEA). Affordable high-performance computing clusters based on commodity hardware and software will soon make large-scale FEA as standard a tool as desktop FEA packages are today. But large-scale FEA has a substantial data management component in defining the problem, building the mesh, running the computation, and analyzing the resulting information. Current, file-based FEA data management practice lacks the scalability and functionality needed to make it easy for engineers to use on huge problems. Introducing Relational Database technology ameliorates both these scalability and usability problems. FEA is not an isolated example of this *size effect*, forcing us to re-think the way we handle application data [1].

This article discusses how to make the transition from a file-based to a database-centric environment in support of large scale FEA. Our reference database system is Microsoft SQL Server 2000 [6,7], but the arguments apply equally well to any RDBMS with good scalability and self-management features. We will point out some of the deficiencies of purely relational systems in the areas of programmability and handling attribute data. That discussion sets the stage for *Part II* which describes the new features and substantial improvements that come with Object-Relational database management systems (ORDBMS) as typified by Microsoft SQL Server 2005 [8,9]. *Part II* shows how these features help address the shortcomings exposed in Part I. *Part III* presents a case study on how client applications can benefit from a database-centric server approach. The sample application is a real-time interactive visualization of three-dimensional metallic polycrystals models using OpenDX [10].

The term Finite Element Analysis (FEA), encompasses the entire process of modeling a physical system, including model generation, meshing, attribute assignment, solution, and post-processing. Each of these phases is a substantial data management task, and the data flows among the phases are also substantial. A general purpose data management solution can accelerate, simplify and unify the entire process not just the individual parts. The methodology presented in these articles has been applied in 3D simulations of crack propagation and in the simulation of plasticity and fatigue of 3D metallic polycrystals in the Cornell Fracture Group [17]. In both cases, there are substantial topology, geometry, and attribute data, and the finite element mesh alone is not sufficient to analyze and interpret the results.

There is an enormous body of recent work on FEA data management [3,13,14,15,16,2,18,19]. Huge advances are being made in each of the six areas outlined in Figure 1. Their scope and implementation varies with the underlying FEA and, in some cases, the intended user interface. Despite the differences in the applications, the recurring themes surrounding the limitations of proprietary file-based FEA data management are the same: the lack of a unified data modeling

> **Figure 1:** FEA Phases
> 1. Model generation
> 2. Meshing
> 3. Attribute assignment
> 4. Solution
> 5. Post Processing
> 6. Visualization Products

capability, the absence of data integrity enforcement and data coherence, limited metadata management, the lack of query interfaces, poor scalability, no support for automatic parallelism, poor application integration, and the lack of support for Internet-scale (Grid [37]) distributed applications. In addressing these issues, some authors have built their own DBMS and query interfaces. We have taken the approach of basing our FEA tools on a supported database system rather than reinvent one.

This paper is *not* an introduction to FEA or to databases. It is at a level that should be accessible to workers in either discipline. We would like to hear your reaction to these ideas, especially if you happen to be an FEA practitioner or a database expert!

## Finite Element Method (FEM)

The Finite Element Method (FEM) is an approach to modeling partial differential equations (PDE) by replacing the continuum problem with an approximate discrete problem suitable for numerical solution in a computer. The discussion here focuses exclusively on common data management problems in the context of FEM-based simulations. No reference is made to and no knowledge of the mathematical theory behind FEM is required [12].

To give a VERY simple example of the FEM approach, imagine that we want to simulate heat flow across a block of inhomogeneous material. The material would be modeled by a space-filling set of voxels (the mesh), each voxel with its own heat conductivity parameters and initial conditions. The differential equations would specify heat transfer among voxels based on the material properties, and the temperature of the neighboring voxels. The solution would iteratively solve these equations occasionally recording the state of the voxels. Post processing would analyze these outputs and put them in a form convenient for visualization.

Other popular discretization methods for PDE include Finite Differences, Finite Volumes, and wavelets. The decision which method to use is driven by the geometric complexity of the underlying domain, the nature of the boundary conditions, as well as the regularity of the PDE. This situation is amplified in multi-physics and multi-scale environments where multiple discretization methods are applied over different but coupled spatial and time scales. Each of these aspects creates specific data modeling and management problems. The conceptual similarity among these problems does not justify automatic application of one solution across the board.

Before we can move on to the core of this paper, the mapping of FEA data onto a relational model, we first sketch the nature, flow, and scale of data in a FEA.

### FEA Data and Dataflow

Various data creation and transformation tasks precede and follow the actual numerical solution of a PDE using FEM. The main phases of a FEA described in Figure 1 are:

*Model creation* generates the topology and geometry information, typically representing the material boundary using parametric surfaces like Bézier patches or Non-Uniform Rational Bézier-Splines (NURBS) [11] in a CAD (Computer Aided Design) tool. The model creation also specifies the system equilibrium and dynamics by defining the interactions as a set of partial differential equations. The *meshing* phase decomposes the model geometry into simple shapes or voxels like tetrahedra or bricks that fill the volume.
*Attribute definition* and *assignment* specifies properties of model entities (e.g., material properties of volumes) and imposes initial and boundary conditions on the solution.
*Solution* uses the model equations, mesh, and attributes to simulate the system's behavior. Often, a whole suite of solutions is run with different attribute assignments. These solutions produce vast quantities of data.
*Post-processing* tasks examine the solution output for features or trends. In our case, tools look for cracks, and quantify crack propagation. During all these phases, visualization tools let the engineers or scientists examine, explore and analyze the data.

This is a simplified presentation of the FEA process. More complex scenarios are quite common: Often the output of one simulation is directly or indirectly part of the input data of other simulations and often the simulation is driven by or interacts with measurements from an experimental system.

Most discussion of FEA focuses on the solution phase. That is were 90% of the data comes from; that is where 90% of the computer time goes. But in our experience, the solution phase consumes a minority of the people-time and people-time is the expensive and time-consuming part of an end-to-end FEA. Although the solution phase produces most of the data, the time spent in the pre-processing and post-processing complex models far exceeds the solution time. As supercomputers and large workstation clusters become ubiquitous, the model preparation and data analysis tasks become even more labor intensive. Computational cluster tools reduce the solution time to several hours or days, and produce vastly more data. On the other hand, creating a correct and complete (CAD) model is a time-consuming and demanding *art*. Analyzing the model outputs is equally demanding and time-consuming.

Reference [5] gives a nice overview of industrial FEA data management. In that presentation, and in general, the terminology is ecumenical -- the term database is used to describe any structured storage beyond a simple file store. The storage spectrum ranges from formatted ASCII or binary files to custom database implementations.

More than 90% of FEA data is easily represented as dense tabular data using basic types like integers and floating point numbers. This includes parts of the topology and geometry, the mesh, and result fields (e.g., displacement, temperature, and stress) produced as part of the solution. But, some of the data, especially attribute data, is more complex.

Reference [4] gives a fairly comprehensive overview of attribute data; both how it is defined and its management requirements. Attributes are used to specify a wide spectrum of FEA parameters, including solution strategy (solver type, tolerances, adaptivity) and data management issues (e.g., sharing and versioning). Although the attribute data is relatively small compared to the solution output, FEA attributes and the associated metadata are fairly complex – involving inheritance from parent structures, and involving complex relationships with neighboring model components and with data from other simulations or from measurements. FEA attributes often have context, scope, and behavior metadata. The attribute's reference-frame is a good example of an attribute context – attributes on curved surfaces are often more easily described using a local non-Cartesian coordinate system, like polar or cylindrical coordinates. An attribute assigned to a volume may be inherited by all voxels in the volume, or by all its bounding faces. Attribute hierarchies and grouping allow sharing and reuse. Time dependent attributes are common. Attributes are also defined as equations that use other attributes as well as references to external resources via hyperlinks, XPointers [33], and WS-Addressing [34].

FEA attributes and metadata are inherently multi-faceted and do not easily lend themselves to a tabular representation. Attributes seem to be well suited to the capabilities of XML and the related X* technologies (XML schema [42], XPath [38], XSLT [41], XQuery [40,39]). This idea is explored later in this series when we talk about SQL Server 2005 features.

We are primarily interested in *unstructured meshes* meaning that the local (graph) structure of the mesh (the number of neighboring vertices) varies from vertex to vertex. Structured or regular meshes allow the data to be stored in a dense array – but adaptive and unstructured meshes require sparse-array or database representation where each voxel or vertex is an independently addressed item. In contrast to a structured mesh, an unstructured mesh is data and metadata at the same time and the metadata cannot be separated from the data.

The entire mesh and most of the attributes serve as input to the FEA solution phase. The mesh is the largest input component - its size depends on (1) the resolution of the discretization, (2) the geometric complexity of the model, and (3) the desired accuracy of output. In practice the mesh varies between hundreds of megabytes to several gigabytes. What's more alarming is that the solution output can be several orders of magnitude larger than the input[1].

In this paper, we are interested in large scale FEA, i.e., meshes with millions of elements resulting in systems of millions of equations. The necessary memory and processing resources needed to obtain a solution *in finite time* are typically not found in workstation class machines. Large multi-processors in the form of clusters of workstations or supercomputers are used instead. Two problems present themselves when working with these large problems and systems:

1. We need an intelligent and scalable (with the size of the objects and the number of nodes) way to move data between the different levels of the memory hierarchy (cluster nodes' memory, local disk, networked file servers, database servers, and tape library). By 'intelligent' we mean that it should be easy to split I/O streams and determine which subset of information needs to go where.

---

[1] For non-linear simulations results are typically stored at the Gauss point level. Each Gauss point hosts a state variable holding $S$ floating point numbers. For $N$ quadratic tetrahedral elements using $G$ Gauss points each with $S$ state variables, the output will include at least $N \cdot S \cdot G$ double precision floating point numbers (8 bytes each). For example, a 3D material stress calculation might have $S = 12$ values for the stress and strain tensors, and $G = 11$, so about $12 \cdot 11 \cdot 8 \sim 1,000$ bytes for each of the $N$ tetrahedral elements. The output consists of many samples of these values during the solution. If $T$ time samples are taken, the solution output is $T \cdot N$ kilobytes.

2. Post-processing typically starts from results in permanent storage. The sheer size asks for analysis in place (including the necessary CPU power) and/or an efficient querying capability.

Apart from size-driven constraints the importance of a simple, data independent definition and manipulation language can hardly be overstated. To fully appreciate this, let's review some basic facts about relational databases.

## Relational Databases

The six phases of Figure 1 are connected by data flows. In general the whole process is data driven. Traditionally, when one changed anything in one phase it had a ripple-effect downstream requiring all the other programs in the pipeline be revised to handle the new data format. As the complexity of FEA systems increases, this tight-coupling between programs and data formats becomes unmentionable.

One of the driving factors behind the development of the relational model was a desire to simplify application design and maintenance. Developers noticed that frequently a substantial part of an (enterprise) application's code was *non problem-oriented code*, code that only parsed, reformatted, or transformed data rather than doing any semantic computation. At the heart of the problem was the poor separation of the physical and logical data layout from the programs [32,1]. Databases solve this problem by introducing the concepts of *schema* and *view*. The schema is a logical and abstract description of the data. A view is a mapping from the schema to a particular data layout as seen by one or more applications. Different programs may want different views. These different views can be automatically created from the same underlying schema and database. In addition, the schema can evolve adding new information, while still providing old programs with the old data views. This *data independence* was a major reason commercial applications adopted database technology. As FEA systems become more complex and more interconnected, this desire for data independence will be a major reason for using database technology in FEA systems.

Current file-oriented FEA systems represent schema metadata as structures (typically in ".h" files.) There is no automatic connection between the files and the corresponding metadata. There are no good tools to evolve the file representation. When one wants to change something, one must change the ".h" file, change the programs, and then reformat the data files. There are self-defining and portable file formats like HDF and NetCDF, but there is no good language support for a view mechanism on them (HDF and NetCDF provide a simple schema mechanism but no view mechanism.) Both are intended for array-oriented data with simple attribute assignment. Several arrays (variables) can be included in a single NetCDF file, but relationships and referential dependence must be stored separately. NetCDF allows constraint-like definitions for variable ranges and extreme values. However, they are not enforced by the standard API. NetCDF is very good at accessing slices or array subsections. Indexing in a wider sense (which would lead to general non-array subsets) is not supported. Write access is limited to one writer. NetCDF can use underlying parallel I/O on some platforms but it is not an inherent part of the API or NetCDF package. In summary, HDF and NetCDF facilities may be adequate for the solution phase of simple structured meshes; but even for those simple problems they do not address the issues of end-to-end analysis because they lack a view mechanism, metadata management, and indexing.

Array-like structures are inadequate for unstructured meshes. Although many mesh constituents can be represented in tabular form, the (irregular) relations between them are an inherent part of the model but cannot be expressed using simple index arithmetic. Most commercial FEA packages supporting unstructured meshes use their own proprietary formats, - creating a myriad of formats/versions and conversion tools. The APIs to manipulate the objects in these files are buried in the corresponding application stacks. As a result, the only safe way to share data is to get a copy in your favored format.

Unlike formatted files, database metadata is unified with the data in a database. Relational databases describe all data in terms of tables. Each table is a set of rows and each row has the same collection of columns. Each column has a name, datatypes, and a set of integrity constraints. Rows in a table are referenced through their identifier or *primary key*, which might consist of one or more columns (compound key). In general, SQL encourages associative and content-based access to data through predicates on the data values. Since for an application the primary key is the only means to reference the data and since applications can be insulated from the data by views, the application will continue to function correctly despite changes in the databases' internal layout and even despite some "external" schema changes – so long as the old data views can be computed from the current data.

The objects in a relational database are defined and manipulated using the Structured Query Language (SQL). SQL is a fairly simple language compared to, say, C++ or even some scripting languages. It is a functional (as opposed to imperative) language, where one specifies what is needed as a *query* -- describing "what is wanted" rather than giving a procedure for "how to get the answer." As other authors have pointed out, coming from an imperative language, the hardest part about learning SQL is to unlearn procedural habits. This means in particular: *Think in terms of sets rather than items, think in terms of goals rather than paths, and think at a high-level!*

Database systems also separate data access from storage representation. One can add indices and reorganize the data as the system evolves without breaking any programs. Indeed, the standard way to speed up a slow program is to add an index to speed its data access. Increasingly, the selection and addition of indices is an automated process – based on the workload.

There are no language extensions or libraries (a la MPI [43] or OpenMP [44]) for "parallel SQL". It is completely up to the optimizer and the database engine to detect and exploit parallelism in a query.

**Microsoft SQL Server 2000 – an Example Relational Database System[2]**

To make this article concrete, we describe our experience with a particular RDBMS, Microsoft SQL Server 2000. The approach described in this paper, can be implemented using almost any database system that has adequate quality, tools, manageability, performance, scalability, security, and integration with the existing infrastructure.

Microsoft SQL Server 2000 [6,7] is available on the Windows platform in various editions. SQL Server's dialect of the SQL language is called Transact-SQL (T-SQL) [22] and is compliant with the entry-level ANSI SQL-92 standard [20]. The built-in types include a variety of integers (1, 2, 4, and 8 bytes long), as well as single and double precision floating point numbers. Fixed and variable-length data types for storing large non-UNICODE and UNICODE character and binary data (like images) are supported.

T-SQL can be used as a programming language for stored procedures and other user-defined functions. User-defined functions are stored with the data in a database and can access it efficiently since they execute as (sandboxed) threads within the SQL process. T-SQL can invoke other processes, send mail, and generally do almost anything (subject to security restrictions), but best-practices argue against this. The standard advice is to just use T-SQL to encapsulate data in the database. External dynamically linked libraries (DLLs) can be accessed from SQL Server through extended stored procedures; but again, this is considered guru-programming and best-practices recommend against it. The general advice is to use SQL as a non-procedural programming language and use it to retrieve exactly the subset of data that you need (retrieve the answers rather than the intermediate results). For example, when a computation is working on a particular region, it should retrieve only the requisite attributes for that region.

SQL Server running on top of Windows supports *symmetric multi-processing* (SMP) architectures (like the 64-way Unisys ES7000 [28]). Effective auto-parallelization is the (un-fulfilled) dream of every MPI or OpenMP programmer. In T-SQL it is reality: Depending on the parallelism inherent in a query and depending on the available resources, the SQL optimizer will generate and schedule multiple tasks (threads, fibers [31]) for a single query. If the hardware resources are adequate, SQL can deliver nearly linear speedup on an SMP. Another means to exploit parallelism in large SQL Server databases are so-called *distributed partitioned views* (DPV) [7]. A logical table (view) might be physically spread across a cluster of multiple linked servers. A query can be directed against any server in the federation, which will determine where the data are located and the query be executed.

Features like high-availability, replication, data mining, and data cubes are supported in SQL Server but are beyond the scope of this discussion.

SQL Server can be accessed from programs running on Windows using the traditional ODBC/JDBC libraries or the more object-oriented OLE/DB and ADO.NET interfaces. UNIX-based systems can interoperate with SQL Server via client APIs like ODBC or OLE/DB.

---

[2] The next version of the product, Microsoft SQL Server 2005, will ship later in 2005. In Part II of this series, we will highlight the new features and extensions beneficial to FEA. Some of the shortcomings with respect to FEA modeling found in SQL Server 2000 have been addressed as well.

## Mapping an FEA Data Model onto a Relational Model

As outlined earlier, the objects in an FEA form a hierarchy implied by the topology geometry, mesh, and attributes that map to physical fields. There are other mappings between the objects of the hierarchy, for example, how a topological face is decomposed into parametric patches and how a given volume is tessellated by tetrahedra. We'll see several examples of complex topology and geometry in Part III that discusses, among other things, the representation of 3D models of metallic polycrystals in a database. Figure 2 gives a hint showing how geometry, material properties, and dynamics interact.

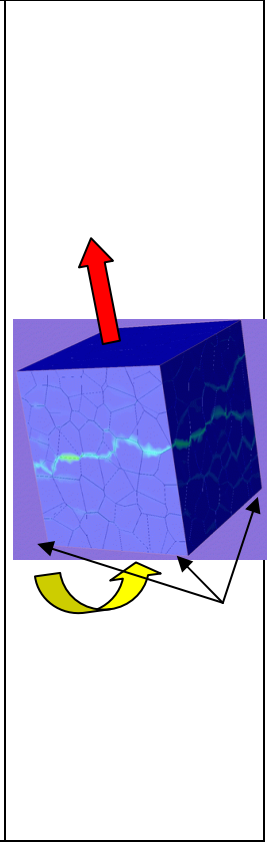

**Figure 2:** A polycrystal model consists of many crystals (grains). Each grain is decomposed into tetrahedra (not shown). Tension is applied at the top surface of the cubical specimen. The bottom face is fixed in normal direction. To prevent rigid body motion, the corners are pinned (the corner closest to the viewer in two directions, the others in just one direction). The material properties assigned to grains and grain boundaries vary from grain to grain and boundary to boundary. The load is applied in increments (and/or cycles), and, after it reaches a certain level, some grain boundaries will start to de-cohere and ultimately fail (de-bond). The underlying database schema has about 60 basic tables plus views and several user-defined functions (see the third article in this series for more details).

Schema design is driven by data semantics and by the questions (queries) we'd like the database to answer. A good starting point for database design is a list of Frequently Asked Queries (FAQ). There is no one FEQ list; different communities have different needs. However, there is a substantial common subset of typical questions among FEA applications. Reference [3] has an example of such a list and its implementation. The list includes simple adjacency queries as well as more advanced queries about geometric proximity and point-in-cell queries. We will see an example of an adjacency query below and come back to geometric queries in the subsection on indexes. Some geometric queries can be seen in action in an animation prepared by Chris Pelkie for the SC2002 conference [2].

The first step in mapping FEA problems into a relational schema requires a mesh representation. Roughly speaking, a mesh or grid is a collection of vertices plus connectivity, in other words, a graph. In addition, there is information about, how a geometric curve is split into mesh edges and the parametric coordinates of the vertices on the curve. The graph structure often exhibits regularity at the local level: Certain clusters of vertices are the corners of the simple shapes or elements (tetrahedra, hexahedra etc.) that form the basis of the mesh.

To translate this to relational database terms, a table of 3D vertices can be defined as follows:

```sql
CREATE TABLE Vertices (
    VertexID int PRIMARY KEY CLUSTERED,
    x float NOT NULL, y float NOT NULL, z float NOT NULL)
```

Although the number of elements sharing a given vertex may vary from vertex to vertex, we know that every tetrahedron has exactly four vertices. Based on this observation, we can define a tetrahedra table as:

```sql
CREATE TABLE Tetrahedra (
    ElemID int NOT NULL PRIMARY KEY,
    v0     int NOT NULL REFERENCES Vertices,
    v1     int NOT NULL REFERENCES Vertices,
    v2     int NOT NULL REFERENCES Vertices,
    v3     int NOT NULL REFERENCES Vertices,
    CONSTRAINT Tetrahedra_UNQ UNIQUE(v0, v1, v2, v3),
    CONSTRAINT Tetrahedra_CHK01 CHECK(v0 != v1),
    -- Add the 5 other pair-wise comparisons
    )
```

This is a very compact definition for tetrahedra. Before we discuss its shortcomings, let's understand its content: Each tetrahedron can be represented as a quadruple of vertex identifiers (its corners). Each tetrahedron vertex ID (v0, v1, v2, v3) must indeed appear as a VertexID in the Vertices table (see the referential constraint). The same quadruple cannot be a representation of different elements (see the uniqueness constraint). A tetrahedron must be non-degenerate, i.e., no two vertex identifiers can be the same (see the six check constraints). For most practical purposes, that is all there is to say about a tetrahedron[3].

The tetrahedra definition above specifies the relation between tetrahedra and vertices. One problem with the definition is that it is not in first-normal form[4] (1NF) [21,23]: Representing the sequence of vertices {v0, v1, v2, v3} as attributes makes some queries awkward. For example, to find all tetrahedra that share a given vertex, the query would be:

```sql
SELECT ElemID
FROM Tetrahedra
WHERE @VertexID in (v0, v1, v2, v3)
```

A normalized version of tetrahedra can be defined as:

```sql
CREATE TABLE TetrahedronVertices (
       ElemID   int      NOT NULL,
       Rank     tinyint  NOT NULL CHECK(Rank < 4),
       VertexID int      NOT NULL REFERENCES Vertices,
       CONSTRAINT PK_TetrahedronVertices PRIMARY KEY(ElemID, Rank),
       CONSTRAINT UNQ_TetrahedronVertices UNIQUE(ElemID, VertexID)
       )
```

This representation expresses the same facts as the previous definition but now the query is more straightforward:

```sql
SELECT ElemID
FROM TetrahedronVertices
WHERE VertexID = @VertexID
```

We could have obtained a normalized view from original table definition as follows:

```sql
CREATE VIEW TetrahedronVertices AS
            SELECT ElemID, 0 AS Rank, v0 AS VertexID FROM Tetrahedra
  UNION ALL SELECT ElemID, 1 AS Rank, v1 AS VertexID FROM Tetrahedra
  UNION ALL SELECT ElemID, 2 AS Rank, v2 AS VertexID FROM Tetrahedra
  UNION ALL SELECT ElemID, 3 AS Rank, v3 AS VertexID FROM Tetrahedra
```

Conversely, we can re-create the quadruple representation from the normalized relation:

```sql
CREATE VIEW Tetrahedra AS
SELECT ElemID,
       A0.VertexID AS v0,
       A1.VertexID AS v1,
       A2.VertexID AS v2,
       A3.VertexID AS v3
  FROM   TetrahedronVertices AS A0
    JOIN TetrahedronVertices AS A1 ON A0.ElemID = A1.ElemID
    JOIN TetrahedronVertices AS A2 ON A1.ElemID = A2.ElemID
    JOIN TetrahedronVertices AS A3 ON A2.ElemID = A3.ElemID
  WHERE A0.Rank = 0 AND A1.Rank = 1 AND A2.Rank = 2 AND A3.Rank = 3
```

What do we do in practice? Either representation works well for small meshes (less than 100,000 elements) – the virtual table (view) lets applications see the other perspective. For large meshes we store both representations and use whichever performs better in a given context.

**FEA Data I/O**

The problems of moving data in and out of databases is often called the *impedance mismatch* – programming language datatypes and procedural interfaces are a different world from tables and non-procedural set-oriented access. There has been a lot of progress in ameliorating the impedance mismatch over the last decade with languages like SQLJ and with interfaces like JDBC and ADO.NET[5]. SQL Server supports several client APIs including ODBC and OLE/DB. Most scripting languages provide convenient libraries or modules for database programming. However, massive data load and dump operations demand the bulk copy client utility (`bcp.exe`), the T-SQL `BULK INSERT` command, or Data Transformation Services (DTS) for better performance (about 10x better).

---

[3] The order of the vertices in a quadruple representing a tetrahedron defines an orientation of the element. However, any even permutation of the vertices defines the same orientation. This fact, for example, is not captured in the above table definition and the insertion of an even (or any non-identical) permutation of a previously inserted element would not be detected as an error. Also the fact that two tetrahedra overlap or that there are unintended gaps would not be detected.

[4] There are five basic normal forms.

The most I/O intensive parts of FEA are:

1. The initial loading of the mesh into the database.
2. The attribute and input data generation for the equation solver.
3. The dumping of analysis results to the database.

Loading the mesh is a straightforward bulk copy and there is nothing particularly interesting or challenging about it. Communicating the mesh and attributes to each of the solver processes is more interesting, and moving the solution output to the database also requires some care.

The solver is an ordinary MPI program running on the cluster of $N$ (~100s) nodes. To run these programs without change, we need to deposit their inputs on the nodes' local disks and harvest their outputs from their local disks. Before the solver runs, each node contacts the database and requests its partition of the mesh. The database responds with the result (a few tens of megabytes) from a simple `JOIN` of the full vertex or voxel table with the partitioning table filtered by partition. (In reality it's slightly more complicated, because neighboring partitions share some voxels or vertexes.) Conversely, when the solver completes (or as it is running) the output data is asynchronously streamed from local file stores into the database.

How do we get those partitions? A "bootstrapping", two-step partitioning approach works well for us. It first generates a few coarse *bootstrap* partitions (typically 8 or 16) in the database using a fast but coarse partitioning method like Recursive Coordinate Bisection [45] written in T-SQL. Then the bootstrap partitions are refined using a high-quality partitioner like ParMetis [26] that, given a finite element graph and the desired number of partitions, maps each vertex and each voxel to one of the N servers. The resulting partitioning is used to scatter the mesh and attributes among the N cluster nodes. (As an optimization, the data is reorganized before being scattered such that the bootstrap nodes communicate only with disjoint node sets.)

Bringing the solution's output data back into the database is decoupled from the job execution on the cluster. The solution output file data is first copied from the nodes onto scratch disks near the database and then bulk-loaded into the database. This is done using the Extract-Transform-Load (ETL) facilities of SQL Server (aka: DTS). It has packages that pull the data from the solvers to a staging area and then bulk load it to the database. It would be a waste of compute resources and a bottleneck if all the nodes attempted a bulk load simultaneously.

Note that SQL Server does not (yet) replace a file server or a parallel file system: All it does is to largely eliminate non problem-oriented code (All it takes to create almost any input format are a few queries!) and store the metadata, like partitioning, with the data. Dedicated I/O systems greatly accelerate the data transfer between file servers and cluster nodes. However, from an end-user's perspective, this does not ease in any way his or her task to extract meaning from the data.

**Indexes**

High-speed query execution is the result of good design and good indexing [23,24]. If the optimizer cannot come up with an efficient execution plan, a query will perform poorly no matter how fast the hardware[6]. Our tables are big (millions of rows) and full table scans are devastating for performance – but an appropriate database index can avoid such scans, taking the query to exactly the desired database subset. Indexes on a table work exactly like the index found at the end of a text book. It is common to implement them as B-trees [23] (B as in *balanced*.). For a *clustered* index, the leaves of the index tree are the rows of the table. For a *non-clustered* index, the leaves are pointers (bookmarks) to actual rows. The SQL query optimizer will automatically discover appropriate indexes and use them when executing the query. FEA geometric queries that determine which cell contains a given point are an easy application of indices. In CAD tools or mesh generators, these queries are usually implemented using octrees. SQL can mimic an octree traversal (this is what the search algorithm does) as an index scan (a B-tree traversal) as follows. Let's assume we are given the following table of non-degenerate cells (boxes):

```sql
CREATE TABLE Cells (
  CellID int PRIMARY KEY,
  x_min float NOT NULL, y_min float NOT NULL, z_min float NOT NULL,
  x_max float NOT NULL, y_max float NOT NULL, z_max float NOT NULL,
  CONSTRAINT CHK_Cells_x CHECK (x_min < x_max),
  CONSTRAINT CHK_Cells_y CHECK (y_min < y_max),
  CONSTRAINT CHK_Cells_z CHECK (z_min < z_max)
)
```

---

[5] The Version 1 of the database interface in the .NET Framework, called ADO.NET, is not intended to handle large datasets and lacks support for bulk operations. ADO.NET 2.0 and SQL Server 2005 address both these issues.

[6] On extremely rare occasions, a query might perform poorly because of server configuration errors or hardware limitations. Windows and SQL Server 2000 ship with excellent diagnostic tools to detect this!

Our point-in-cell query would look something like this:

```sql
SELECT CellID FROM Cells
  WHERE @x BETWEEN x_min AND x_max
    AND @y BETWEEN y_min AND y_max
    AND @z BETWEEN z_min AND z_max
```

Without an index, a full table scan will be performed. A simple index that will considerably reduce the search space is:

```sql
CREATE INDEX Idx_Cells ON Cells (x_min, x_max, y_min, y_max)
```

Detecting the index, the optimizer will translate the query into an index seek plus bookmark lookup[7]. More sophisticated geometric queries can be defined -- after all, not everything is a box! Also, space-filling curve schemes (Peano, Hilbert curves and others [46]) can be used to generate surrogate keys. Often fast lookups combine a cheap, coarse-grained search producing a small candidate list with an (expensive) brute-force search on a small subset. We use all these techniques to achieve good spatial indexes [27].

**FEA Attributes and Metadata**

So far the discussion focused on dense tabular data which can be easily incorporated in a relational model. It is true that the bulk of FEA data fits this model; but our solution would be of limited use if we couldn't integrate the remaining 1% of mostly attributes and metadata. After all, the main argument for using a database is to unify the data and meta-data and to provide a common store for both that all parts of the FEA can use.

References [4,5] explain that attributes come in many forms: boundary conditions, material properties, solver parameters, control information and so on. In most cases, it would be possible to force them into a tabular form, but that would be a clumsy and brittle representation, and, for the lack of manipulative power, the management API would have to be implemented outside the database. This is another example of the *impedance mismatch* between traditional database models and real applications. XML and the related X* technologies (XML schema, XPath, XSLT, XQuery etc.) are ideal candidates to bridge this gap. SQL Server 2000 support for XML [30] is limited to conversion from relational data to XML (`FOR XML`, XML templates) and the extraction of relational data from XML documents (`OPENXML`). Put another way, XML is an add-on to SQL Server 2000 rather than being a fully integrated datatype. The newer version of SQL Server (2005) and newer versions of other systems like DB2 and Oracle have much better integration and support for XML. Part II of this series shows how we have used XML and SQL Server 2005 to solve our problems! Before that, we used the `text` data type to store ASCII representations of XML documents in a table and did the XML processing using the .NET Framework on the client side. The `image` data type was certainly adequate for storing binary data, including manually serialized objects.

This brings us to the question of user-defined types (UDT) in general. "A user-defined datatype provides a convenient way for you to guarantee consistent use of underlying native datatypes for columns known to have the same domain of possible values." ([7], page 239) This is an interesting way of saying that there are no 'real' UDTs in SQL Server 2000[8]. In fact, there are none in the relational model and they are against the spirit of normalization and non-procedural programming. The SQL-99 standard introduced collection types like `ARRAY` and `ROW` which would allow arrays and C-struct like UDTs. In Part II, we will see that almost any public class of the .NET Common Language Runtime (CLR) can be used as a UDT in SQL Server 2005. This does not mean that the tabular structure will be submerged in or replaced by the UDT. To get all the SQL performance benefits (sorting, indexing) and not overly suffer from CLR invocation overhead, a well thought out table and UDT co-design are necessary (see Part II).

---

[7] If we had defined the index to be a clustered index, no bookmark lookup would be necessary. In SQL Server, the number of clustered indexes per table is limited to one.

[8] A heroic effort to introduce complex arithmetic on SQL Server 2000 can be found in [29].

**Practical Considerations**

This subsection gives some practical advice about using SQL Server 2000 for FEA. The Microsoft SQL Server 2000 Desktop Engine (MSDE 2000) is available as a free download from Microsoft [35]. If you are just starting, make sure you start with SQL Server 2005. The tools and basic functionality are identical to the enterprise editions – but it is limited to storing 1 GB databases. For academic users, the MSDN Academic Alliance gives you (nearly) free access to all the development tools and server products, including SQL Server and the programming tools. The client tools of all SQL Server editions include the Query Analyzer, Enterprise Manager, Online Books, as well as a collection of command line tools. The Query Analyzer is a graphical shell which can be used for query execution and debugging, as well as for analysis and profiling of execution plans. The Enterprise Manager is the main administrative tool. It can also be used as a design tool and to generate diagrams of database schemas. It is helpful for SQL novices to use a graphical designer, but we recommend you take control and become fluent in T-SQL.

Databases grow automatically, but you may not want to unconditionally rely on auto-growth. The default growth rate is 10% and, depending on your needs, you might want to set it to something more generous, or manually adjust the size.

SQL Server 2000 does many good things for you without you knowing about it, in some cases more than you need. After all, it was designed to perform well in an enterprise environment, doing online transaction processing (OLTP) or data warehousing. Naturally, it can do many things we do not or cannot take advantage of that may get in our way. For example, by default SQL Server keeps a log of all your changes until you archive the log. This enables you to go back in time to pretty much any previous state of the database. (Version control systems for files are not quite as good, but similar.) This luxury does not come for free, i.e. without a performance impact, but it is the (correct) default behavior. For the data access and update patterns in FEA described in this paper a more relaxed *simple recovery model* can be applied [7].

Column and table constraints reflect important assumptions about the data. An FEA database typically has what we would call primary and derived tables. The former are populated through bulk operations from raw (in general, unchecked!) data. Checking constraints at this level is of particular importance[9]. The derived tables are generated based on the primary tables and the validity of the imposed primary constraints is assumed. For performance (and after testing!), you might want to keep the constraints on secondary tables to a minimum. Otherwise, like it or not, on any `INSERT SELECT`, `UPDATE` etc. SQL Server will check these constraints and the derived table generation (if they are large) might be slower than expected. During insert operations you can aggressively lock these tables using, for example, the `TABLOCK` hint.

As far as hardware goes, you should get as much memory and fast a disk subsystem as you can afford (CPU speed is less of an issue than I/O and memory)[10]. To start, just spread all the data across all the disks, but if you have performance problems segregate log files to different drives. If the hardware is sufficiently reliable (if you can tolerate reloading or regenerating the data), use RAID 0 (striped disks) to maximize I/O performance.

Though not formally ANSI/IEEE 754 [36] compliant, SQL Server 2000 does a good job for floating point arithmetic and conversion of string representations of floating point numbers, as long as you stay within the limits set out by the standard. Caution is advised, if your application "relies" on standard conforming exception handling: The SQL Standard and SQL Server 2000 have no notion of NaN.

Normalization [21,22] is an important concept in relational design. The goal is to eliminate redundancy in the representation of facts, which otherwise might lead to inconsistent updates. (Since the facts to be modeled determine normality, it is inaccurate to speak of the normality of tables independent of the underlying facts.) Performance might sometimes overrule normalization, but non-redundant representation should be the starting point of (and will ease!) your design. As in "real" code, avoid pre-mature optimization!

---

[9] Beware that the default behavior of bulk operations is to disable constraint checking (for performance). You can either check the constraints manually, with a query, after the bulk operation (recommended) or force the checking during bulk loading (not recommended).

[10] Indexes certainly work no matter if the data pages are resident in memory or on disk. We have worked with databases of 2 TB on machines with 4 GB of RAM. Of course, the more you can hold in main memory the better.

Finally, don't be fooled by the almost exclusive coverage of business related applications in the (vast) database literature. Learning the basic moves is not very difficult, but we acknowledge that, for the time being, the supply of nutritious science or engineering related examples is rather limited. You can help change this!

## Final Remarks

This first part in the series describes how our FEA pipeline uses SQL Server as a common data store for both the simulation data and the metadata. It uses SQL Server as a data-shell, as the communication medium and glue among the six solution phases (modeling, meshing, attribute generation, solution, post-processing, and visualization). Subsequent articles will describe how the design has evolved to leverage the new object-oriented and XML features of Object-Relational database systems.

Though it might seem straightforward, it is not easy to quantify the advantages of the approach advocated in this article over the traditional approach to FEA data management. Metrics suggesting themselves are development time, code size, maintainability, administrative and infrastructure cost, and turnaround. Among the implicit assumptions would be a fixed problem size and complexity, equivalent task structure, and the same underlying hardware. In practice, this is hardly ever (read never) the case: In our own simulations, we have seen the problem sizes grow by more than two orders of magnitude over the last five years. The hardware to accommodate these problems grew proportionally. The search for data management alternatives was first and foremost driven by this (continuing) drastic increase in problem size and complexity, - not by the desire to write a paper. (In that sense, we threw away the ladder for something better, after we climbed up on it[11].)

There are many moving parts in an FEA: Geometric modelers, mesh generators, attribute managers, and visualization tools are under constant development, and new requirements are emerging (Web services). This development should be as concurrent and decoupled as possible. The individual components deal with different subsets of the data, but must not be locked into a particular view or subset because requirements might change. The fundamental data management question is how we can efficiently define, debug, diagnose, verify, select, manipulate and transform data, and ultimately *guarantee* the sanity and integrity of the data. Without a well-defined data model these terms don't mean anything and separation of the metadata from the data has to be paid for, among other things, with replicated and brittle non problem-oriented code. If we had to summarize the key lesson learned in one sentence, it would read like this: *Using a database-centric approach puts you in control of your data regardless of their size.*

## Acknowledgements

We gratefully acknowledge the support of Microsoft and Microsoft Research. The first author enjoyed the support of the Cornell Theory Center, and of the head of the Cornell Fracture Group, Professor Anthony R. Ingraffea. He would like to thank all members of the Cornell Fracture Group for years of inspiring collaborations and for their continued help in understanding the challenges of engineering.

---

[11] Ludwig Wittgenstein, Tractatus Logico-Philosophicus, 6.54